# THE VERDOORN LAW IN THE PORTUGUESE REGIONS: A PANEL DATA ANALYSIS

**Vitor João Pereira Domingues Martinho**

Unidade de I&D do Instituto Politécnico de Viseu
Av. Cor. José Maria Vale de Andrade
Campus Politécnico
3504 - 510 Viseu
**(PORTUGAL)**
**e-mail:** vdmartinho@esav.ipv.pt

**ABSTRACT**

This work aims to test the Verdoorn Law, with the alternative specifications of (1)Kaldor (1966), for five regions (NUTS II) Portuguese from 1986 to 1994 and for the 28 NUTS III Portuguese in the period 1995 to 1999. Will, therefore, to analyze the existence of increasing returns to scale that characterize the phenomena of polarization with circular and cumulative causes and can explain the processes of regional divergence. It is intended to test, even in this work, the alternative interpretation of (2)Rowthorn (1975) Verdoorn's Law for the same regions and periods. The results of this work will be complemented with estimates of these relationships to other sectors of the economy than the industry (primary and services sector), for each of the manufacturing industries operating in the Portuguese regions and for the total economy of each region (3)(Martinho, 2011).

## 1. INTRODUCTION

(4)Verdoorn (1949) was the first author to reveal the importance of the positive relationship between the growth of labor productivity and output growth, arguing that the causality is from output to productivity, thus assuming that labor productivity is endogenous. An important finding of the empirical relationship is the elasticity of labor productivity with respect to output that according to Verdoorn is approximately 0.45 on average, external limits between 0.41 and 0.57. This author also found that the relationship between productivity growth and output growth reflects a kind of production technology and the existence of increasing returns to scale, which contradicts the hypothesis of neoclassical constant returns to scale, or decreasing, and absolute convergence Regional.

Kaldor rediscovered this law in 1966 and since then Verdoorn's Law has been tested in several ways, using specifications, samples and different periods. However, the conclusions drawn differ, some of them rejecting the Law of Verdoorn and other supporting its validity. (5)Kaldor (1966, 1967) in his attempt to explain the causes of the low rate of growth in the UK, reconsidering and empirically investigating Verdoorn's Law, found that there is a strong positive relationship between the growth of labor productivity (p) and output (q), i.e. p = f (q). Or alternatively between employment growth (e) and the growth of output, ie, e = f (q).

Another interpretation of Verdoorn's Law, as an alternative to the Kaldor, is presented by (6)Rowthorn (1975, 1979). Rowthorn argues that the most appropriate specification of Verdoorn's Law is the ratio of growth of output (q) and the growth of labor productivity (p) with employment growth (e), i.e., q = f (e) and p = f (e), respectively (as noted above, the exogenous variable in this case is employment). On the other hand, Rowthorn believes that the empirical work of Kaldor (1966) for the period 1953-54 to 1963-64 and the (7)Cripps and Tarling (1973) for the period 1951 to 1965 that confirm Kaldor's Law, not can be accepted since they are based on small samples of countries, where extreme cases end up like Japan have great influence on overall results.

It should be noted, finally, that several authors have developed a body of work in order to test the Verdoorn's Law in a regional context, including (8)Leon-Ledesma (1998).

## 2. ALTERNATIVE SPECIFICATIONS OF VERDOORN'S LAW

The hypothesis of increasing returns to scale in industry was initially tested by Kaldor (1966) using the following relations:

$p_i = a + bq_i$, Verdoorn law (1)

$e_i = c + dq_i$, Kaldor law (2)

where pi, qi and ei are the growth rates of labor productivity, output and employment in the industrial sector in the economy i.

On the other hand, the mathematical form of Rowthorn specification is as follows:

$$p_i = \lambda_1 + \varepsilon_1 e_i, \text{ firts equation of Rowthorn (3)}$$

$$q_i = \lambda_2 + \varepsilon_2 e_i, \text{ second equation of Rowthorn (4)}$$

where $\lambda_1 = \lambda_2$ e $\varepsilon_2 = (1+\varepsilon_1)$, because $p_i = q_i - e_i$. In other words, $q_i - e_i = \lambda_1 + \varepsilon_1 e_i$, $q_i = \lambda_1 + e_i + \varepsilon_1 e_i$, so, $q_i = \lambda_1 + (1+\varepsilon_1)e_i$.

Rowthorn estimated these equations for the same OECD countries considered by Kaldor (1966), with the exception of Japan, and for the same period and found that $\varepsilon_2$ was not statistically different from unity and therefore $\varepsilon_1$ was not statistically different from zero. This author thus confirmed the hypothesis of constant returns to scale in manufacturing in the developed countries of the OECD. (9)Thirlwall (1980) criticized these results, considering that the Rowthorn interpretation of Verdoorn's Law is static, since it assumes that the Verdoorn coefficient depends solely on the partial elasticity of output with respect to employment.

### 3. DATA ANALYSIS

Considering the variables on the models of Kaldor and Rowthorn presented previously and the availability of statistical information, we used the following data disaggregated at regional level. Annual data for the period 1986 to 1994 corresponding to the five regions of mainland Portugal (NUTS II) for the different economic sectors, including the various manufacturing industries in those regions and the total economy of these regions. These data were obtained from Eurostat (Eurostat Regio of Statistics 2000). We also used data for the period from 1995 to 1999 of the twenty-eight NUTS III regions of mainland Portugal and with the same sectoral breakdown mentioned above. The data for the period 1995 to 1999 were obtained from the INE (National Accounts 2003).

### 4. EMPIRICAL EVIDENCE OF THE VERDOORN'S LAW

The results in Table 1, obtained in the estimations carried out with the equations of Verdoorn, Kaldor and Rowthorn for each of the sectors of the economy and for the total economy of each of the five regions considered in the first period, to state the following.

The industry is the sector that has the biggest increasing returns to scale, followed by agriculture and service sector. Services without the public sector present values for the income scale unacceptable and manufacturing presents surprisingly very low values, reflecting a more intensive use of labor.

It should be noted, finally, for this set of results the following table: Verdoorn's equation is the most satisfactory in terms of statistical significance of the coefficient obtained and the degree of explanation in the various estimations. There is, therefore, that productivity is endogenous and generated by the growth of regional and sectoral output.

**Table 1:** Analysis of economies of scale through the equation Verdoorn, Kaldor and Rowthorn, for each of the economic sectors and the five NUTS II of Portugal, for the period 1986 to 1994

| Agriculture | | | | | | |
|---|---|---|---|---|---|---|
| | Constant | Coefficient | DW | $R^2$ | G.L. | E.E. (1/(1-b)) |
| **Verdoorn** $p_i = a + bq_i$ | 0.042* (5.925) | 0.878* (12.527) | 1.696 | 0.805 | 38 | 8.197 |
| **Kaldor** $e_i = c + dq_i$ | -0.042* (-5.925) | 0.123** (1.750) | 1.696 | 0.075 | 38 | |
| **Rowthorn1** $p_i = \lambda_1 + \varepsilon_1 e_i$ | -0.010 (-0.616) | -0.621** (-1.904) | 1.568 | 0.087 | 38 | |
| **Rowthorn2** $q_i = \lambda_2 + \varepsilon_2 e_i$ | -0.010 (-0.616) | 0.379 (1.160) | 1.568 | 0.034 | 38 | |
| **Industry** | | | | | | |
| | Constant | Coefficient | DW | $R^2$ | G.L. | E.E. (1/(1-b)) |
| **Verdoorn** | -12.725* (-4.222) | 0.992* (8.299) | 2.001 | 0.587 | 37 | 125.000 |
| **Kaldor** | 12.725* (4.222) | 0.008 (0.064) | 2.001 | 0.869 | 37 | |
| **Rowthorn1** | 15.346* (9.052) | -0.449* (-3.214) | 1.889 | 0.326 | 37 | |
| **Rowthorn2** | 15.346* (9.052) | 0.551* (3.940) | 1.889 | 0.776 | 37 | |

| Manufactured Industry | | | | | | |
|---|---|---|---|---|---|---|
| | Constant | Coefficient | DW | $R^2$ | G.L. | E.E. (1/(1-b)) |
| Verdoorn | 8.296* (4.306) | 0.319* (2.240) | 1.679 | 0.139 | 37 | 1.468 |
| Kaldor | -8.296* (-4.306) | 0.681* (4.777) | 1.679 | 0.887 | 37 | |
| Rowthorn1 | 12.522* (12.537) | -0.240* (-2.834) | 1.842 | 0.269 | 37 | |
| Rowthorn2 | 12.522* (12.537) | 0.760* (8.993) | 1.842 | 0.891 | 37 | |
| Services | | | | | | |
| | Constant | Coefficient | DW | $R^2$ | G.L. | E.E. (1/(1-b)) |
| Verdoorn | -0.045* (-3.253) | 0.802* (6.239) | 1.728 | 0.506 | 38 | 5.051 |
| Kaldor | 0.045* (3.253) | 0.198 (1.544) | 1.728 | 0.059 | 38 | |
| Rowthorn1 | 0.071* (4.728) | -0.694* (-3.607) | 1.817 | 0.255 | 38 | |
| Rowthorn2 | 0.071* (4.728) | 0.306 (1.592) | 1.817 | 0.063 | 38 | |
| Services (without public sector) | | | | | | |
| | Constant | Coefficient | DW | $R^2$ | G.L. | E.E. (1/(1-b)) |
| Verdoorn | -0.074* (-4.250) | 1.020* (7.695) | 1.786 | 0.609 | 38 | --- |
| Kaldor | 0.074* (4.250) | -0.020 (-0.149) | 1.786 | 0.001 | 38 | |
| Rowthorn1 | 0.076* (4.350) | -0.903* (-4.736) | 1.847 | 0.371 | 38 | |
| Rowthorn2 | 0.076* (4.350) | 0.097 (0.509) | 1.847 | 0.007 | 38 | |
| All Sectors | | | | | | |
| | Constant | Coefficient | DW | $R^2$ | G.L. | E.E. (1/(1-b)) |
| Verdoorn | -0.020* (-2.090) | 0.907* (8.367) | 1.595 | 0.648 | 38 | 10.753 |
| Kaldor | 0.020* (2.090) | 0.093 (0.856) | 1.595 | 0.019 | 38 | |
| Rowthorn1 | 0.056* (6.043) | -0.648* (-2.670) | 2.336 | 0.255 | 32 | |
| Rowthorn2 | 0.056* (6.043) | 0.352 (1.453) | 2.336 | 0.225 | 32 | |

**Note: * Coefficient statistically significant at 5%, ** Coefficient statistically significant at 10%, GL, Degrees of freedom; EE, Economies of scale.**

Applying the same methodology for each of the manufacturing industries, we obtained the results presented in Table 2.

Manufacturing industries that have, respectively, higher increasing returns to scale are the industry of transport equipment (5.525), the food industry (4.274), industrial minerals (3.906), the metal industry (3.257), the several industry (2.222), the textile industry (1.770), the chemical industry (1.718) and industry equipment and electrical goods (presents unacceptable values). The paper industry has excessively high values. Note that, as expected, the transportation equipment industry and the food industry have the best economies of scale (they are modernized industries) and the textile industry has the lowest economies of scale (industry still very traditional, labor intensive, and in small units).

Also in Table 2 presents the results of an estimation carried out with 9 manufacturing industries disaggregated and together (with 405 observations). By analyzing these data it appears that were obtained respectively for the coefficients of the four equations, the following elasticities: 0.608, 0.392, -0.275 and 0.725. Therefore, values that do not indicate very strong increasing returns to scale, as in previous estimates, but are close to those obtained by Verdoorn and Kaldor.

**Table 2:** Analysis of economies of scale through the equation Verdoorn, Kaldor and Rowthorn, for each of the manufacturing industries and in the five NUTS II of Portugal, for the period 1986 to 1994

| Metal Industry | | | | | | |
|---|---|---|---|---|---|---|
| | Constant | Coefficient | DW | $R^2$ | G.L. | E.E. (1/(1-b)) |
| Verdoorn $p_i = a + bq_i$ | -4.019* (-2.502) | 0.693* (9.915) | 1.955 | 0.898 | 29 | 3.257 |

| | Kaldor $e_i = c + dq_i$ | 4.019*<br>(2.502) | 0.307*<br>(4.385) | 1.955 | 0.788 | 29 | |
|---|---|---|---|---|---|---|---|
| | Rowthorn1 $p_i = \lambda_1 + \varepsilon_1 e_i$ | -12.019<br>(-0.549) | 0.357<br>(1.284) | 1.798 | 0.730 | 29 | |
| | Rowthorn2 $q_i = \lambda_2 + \varepsilon_2 e_i$ | -12.019<br>(-0.549) | 1.357*<br>(4.879) | 1.798 | 0.751 | 29 | |

**Mineral Industry**

| | Constant | Coefficient | DW | $R^2$ | G.L. | E.E. (1/(1-b)) |
|---|---|---|---|---|---|---|
| Verdoorn | -0.056*<br>(-4.296) | 0.744*<br>(4.545) | 1.978 | 0.352 | 38 | 3.906 |
| Kaldor | 0.056*<br>(4.296) | 0.256<br>(1.566) | 1.978 | 0.061 | 38 | |
| Rowthorn1 | -0.023<br>(-0.685) | -0.898*<br>(-9.503) | 2.352 | 0.704 | 38 | |
| Rowthorn2 | -0.023<br>(-0.685) | 0.102<br>(1.075) | 2.352 | 0.030 | 38 | |

**Chemical Industry**

| | Constant | Coefficient | DW | $R^2$ | G.L. | E.E. (1/(1-b)) |
|---|---|---|---|---|---|---|
| Verdoorn | 0.002<br>(0.127) | 0.418*<br>(6.502) | 1.825 | 0.554 | 34 | 1.718 |
| Kaldor | -0.002<br>(-0.127) | 0.582*<br>(9.052) | 1.825 | 0.707 | 34 | |
| Rowthorn1 | 9.413*<br>(9.884) | 0.109<br>(0.999) | 1.857 | 0.235 | 33 | |
| Rowthorn2 | 9.413*<br>(9.884) | 1.109*<br>(10.182) | 1.857 | 0.868 | 33 | |

**Electrical Industry**

| | Constant | Coefficient | DW | $R^2$ | G.L. | E.E. (1/(1-b)) |
|---|---|---|---|---|---|---|
| Verdoorn | 0.004<br>(0.208) | -0.126<br>(-1.274) | 1.762 | 0.128 | 32 | --- |
| Kaldor | -0.004<br>(-0.208) | 1.126*<br>(11.418) | 1.762 | 0.796 | 32 | |
| Rowthorn1 | 0.019<br>(1.379) | -0.287*<br>(-4.593) | 1.659 | 0.452 | 32 | |
| Rowthorn2 | 0.019<br>(1.379) | 0.713*<br>(11.404) | 1.659 | 0.795 | 32 | |

**Transport Industry**

| | Constant | Coefficient | DW | $R^2$ | G.L. | E.E. (1/(1-b)) |
|---|---|---|---|---|---|---|
| Verdoorn | -0.055*<br>(-2.595) | 0.819*<br>(5.644) | 2.006 | 0.456 | 38 | 5.525 |
| Kaldor | 0.055*<br>(2.595) | 0.181<br>(1.251) | 2.006 | 0.040 | 38 | |
| Rowthorn1 | -0.001<br>(-0.029) | -0.628*<br>(-3.938) | 2.120 | 0.436 | 32 | |
| Rowthorn2 | -0.001<br>(-0.029) | 0.372*<br>(2.336) | 2.120 | 0.156 | 32 | |

**Food Industry**

| | Constant | Coefficient | DW | $R^2$ | G.L. | E.E. (1/(1-b)) |
|---|---|---|---|---|---|---|
| Verdoorn | 0.006<br>(0.692) | 0.766*<br>(6.497) | 2.191 | 0.526 | 38 | 4.274 |
| Kaldor | -0.006<br>(-0.692) | 0.234**<br>(1.984) | 2.191 | 0.094 | 38 | |
| Rowthorn1 | 0.048*<br>(2.591) | -0.679*<br>(-4.266) | 1.704 | 0.324 | 38 | |
| Rowthorn2 | 0.048*<br>(2.591) | 0.321*<br>(2.018) | 1.704 | 0.097 | 38 | |

**Textile Industry**

| | Constant | Coefficient | DW | $R^2$ | G.L. | E.E. (1/(1-b)) |
|---|---|---|---|---|---|---|
| Verdoorn | -0.008<br>(-0.466) | 0.435*<br>(3.557) | 2.117 | 0.271 | 34 | 1.770 |
| Kaldor | 0.008<br>(0.466) | 0.565*<br>(4.626) | 2.117 | 0.386 | 34 | |
| Rowthorn1 | 0.002<br>(0.064) | -0.303*<br>(-2.311) | 1.937 | 0.136 | 34 | |
| Rowthorn2 | 0.002 (0.064) | 0.697* (5.318) | 1.937 | 0.454 | 34 | |
| **Paper Industry** | | | | | | |
| | **Constant** | **Coefficient** | **DW** | **R²** | **G.L.** | **E.E. (1/(1-b))** |
| Verdoorn | -0.062* (-3.981) | 1.114* (12.172) | 1.837 | 0.796 | 38 | ∞ |
| Kaldor | 0.062* (3.981) | -0.114 (-1.249) | 1.837 | 0.039 | 38 | |
| Rowthorn1 | 0.028 (1.377) | -1.053* (-4.134) | 1.637 | 0.310 | 38 | |
| Rowthorn2 | 0.028 (1.377) | -0.053 (-0.208) | 1.637 | 0.001 | 38 | |
| **Several Industry** | | | | | | |
| | **Constant** | **Coefficient** | **DW** | **R²** | **G.L.** | **E.E. (1/(1-b))** |
| Verdoorn | -1.212 (-0.756) | 0.550* (8.168) | 2.185 | 0.529 | 37 | 2.222 |
| Kaldor | 1.212 (0.756) | 0.450* (6.693) | 2.185 | 0.983 | 37 | |
| Rowthorn1 | 8.483* (24.757) | 0.069 (1.878) | 2.034 | 0.175 | 37 | |
| Rowthorn2 | 8.483* (24.757) | 1.069* (29.070) | 2.034 | 0.975 | 37 | |
| **9 Manufactured Industry Together** | | | | | | |
| | **Constant** | **Coefficient** | **DW** | **R²** | **G.L.** | **E.E. (1/(1-b))** |
| Verdoorn | -0.030* (-6.413) | 0.608* (19.101) | 1.831 | 0.516 | 342 | 2.551 |
| Kaldor | 0.030* (6.413) | 0.392* (12.335) | 1.831 | 0.308 | 342 | |
| Rowthorn1 | -0.003 (-0.257) | -0.275* (-4.377) | 1.968 | 0.053 | 342 | |
| Rowthorn2 | -0.003 (-0.257) | 0.725* (11.526) | 1.968 | 0.280 | 342 | |

**Note: * Coefficient statistically significant at 5%, ** Coefficient statistically significant at 10%, GL, Degrees of freedom; EE, Economies of scale.**

At Table 3, with results of estimations performed for each of the sectors and in the period 1995 to 1999, to stress again that the industry has the greatest increasing returns to scale (9.091), followed by services (1.996). Agriculture, in turn, presents unacceptable values.

In Table 4 are the results of an estimation carried out for nine manufacturing industries disaggregated and together, as in the face of data availability (short period of time and lack of disaggregated data for these industries in NUTS III) this is a way to estimate considered the equations for the different manufacturing industries during this period. For the analysis of the data reveals that the values of the coefficients of the four equations are, respectively, 0.774, 0.226, -0.391 and 0.609 (all statistically significant), reflecting the increasing returns to scale increased slightly in this economic sector, i.e. of 2.551 (Table 2) to 4.425.

**Table 3:** Analysis of economies of scale through the equation Verdoorn, Kaldor and Rowthorn, for each of the economic sectors and NUTS III of Portugal, for the period 1995 to 1999

| **Agriculture** | | | | | | |
|---|---|---|---|---|---|---|
| | **Constant** | **Coefficient** | **DW** | **R²** | **G.L.** | **E.E. (1/(1-b))** |
| Verdoorn[(1)] | 0.010 (0.282) | 0.053 (0.667) | 0.542 | 1.690 | 23 | --- |
| Verdoorn $p_i = a + bq_i$ | 0.023* (3.613) | 1.105* (17.910) | 1.959 | 0.745 | 110 | |
| Kaldor $e_i = c + dq_i$ | -0.023* (-3.613) | -0.105** (-1.707) | 1.959 | 0.026 | 110 | |
| Rowthorn1 $p_i = \lambda_1 + \varepsilon_1 e_i$ | -0.032* (-5.768) | -1.178* (-9.524) | 1.713 | 0.452 | 110 | |
| Rowthorn2 $q_i = \lambda_2 + \varepsilon_2 e_i$ | -0.032* (-5.768) | -0.178 (-1.441) | 1.713 | 0.019 | 110 | |
| **Industry** | | | | | | |
| | **Constant** | **Coefficient** | **DW** | **R²** | **G.L.** | **E.E. (1/(1-b))** |

| | Constant | Coefficient | DW | R² | G.L. | E.E. (1/(1-b)) |
|---|---|---|---|---|---|---|
| Verdoorn[1] | 0.017 (0.319) | 0.053 (0.673) | 0.195 | 2.380 | 23 | |
| Verdoorn | -0.014* (-2.993) | 0.890* (18.138) | 2.253 | 0.749 | 110 | |
| Kaldor | 0.014* (2.993) | 0.110* (2.236) | 2.253 | 0.044 | 110 | 9.091 |
| Rowthorn1 | 0.053* (6.739) | -0.617* (-3.481) | 2.069 | 0.099 | 110 | |
| Rowthorn2 | 0.053* (6.739) | 0.383* (2.162) | 2.069 | 0.041 | 110 | |
| **Services** | | | | | | |
| | Constant | Coefficient | DW | R² | G.L. | E.E. (1/(1-b)) |
| Verdoorn[1] | 0.003 (0.306) | 0.096* (8.009) | 0.773 | 2.492 | 23 | |
| Verdoorn | 0.007 (1.098) | 0.499* (6.362) | 2.046 | 0.269 | 110 | |
| Kaldor | -0.007 (-1.098) | 0.502* (6.399) | 2.046 | 0.271 | 110 | 1.996 |
| Rowthorn1 | 0.059* (19.382) | -0.432* (-5.254) | 1.993 | 0.201 | 110 | |
| Rowthorn2 | 0.059* (19.382) | 0.568* (6.895) | 1.993 | 0.302 | 110 | |
| **All Sectors** | | | | | | |
| | Constant | Coefficient | DW | R² | G.L. | E.E. (1/(1-b)) |
| Verdoorn[1] | 0.007 (0.188) | 0.090* (2.524) | 0.203 | 2.588 | 23 | |
| Verdoorn | -0.015* (-3.245) | 0.851* (13.151) | 2.185 | 0.611 | 110 | |
| Kaldor | 0.015* (3.245) | 0.149* (2.308) | 2.185 | 0.046 | 110 | 6.711 |
| Rowthorn1 | 0.057* (13.017) | -0.734* (-5.499) | 2.092 | 0.216 | 110 | |
| Rowthorn2 | 0.057* (13.017) | 0.266** (1.989) | 2.092 | 0.035 | 110 | |

**Note: (1) cross-section Estimation * Coefficient statistically significant at 5%, ** Coefficient statistically significant at 10%, GL, Degrees of freedom; EE, Economies of scale.**

**Table 4:** Analysis of economies of scale through the equation Verdoorn, Kaldor and Rowthorn, for nine manufacturing industries together for the period 1995 to 1999 and five in mainland Portugal NUTS II

| 9 Manufactured Industry Together | | | | | | |
|---|---|---|---|---|---|---|
| | Constant | Coefficient | DW | R² | G.L. | E.E. (1/(1-b)) |
| Verdoorn $p_i = a + bq_i$ | 0.004 (0.766) | 0.774* (20.545) | 2.132 | 0.703 | 178 | |
| Kaldor $e_i = c + dq_i$ | -0.004 (-0.766) | 0.226* (6.010) | 2.132 | 0.169 | 178 | 4.425 |
| Rowthorn1 $p_i = \lambda_1 + \varepsilon_1 e_i$ | 0.049* (4.023) | -0.391* (-3.392) | 2.045 | 0.112 | 132 | |
| Rowthorn2 $q_i = \lambda_2 + \varepsilon_2 e_i$ | 0.049* (4.023) | 0.609* (5.278) | 2.045 | 0.214 | 132 | |

**Note: * Coefficient statistically significant at 5%, ** Coefficient statistically significant at 10%, GL, Degrees of freedom; EE, Economies of scale.**

## 5. CONCLUSIONS

In the estimates made for each of the economic sectors in the first period (1986-1994), it appears that the industry is the largest that has increasing returns to scale, followed by agriculture and service sector.

At the level of estimates made for manufacturing industries, it appears that those with, respectively, higher yields are industry transport equipment, food industry, industrial minerals, metals industry, the several industries, the textile industry, chemical industry and industry equipment and electrical goods. The paper industry has excessively high values.

The results of the estimations made for each of the economic sectors in the second period (1995-1999), notes that the industry again provides greater increasing returns to scale, followed by services. Agriculture, on the other hand, has overly high values.